\documentclass{article}

\usepackage{microtype}
\usepackage{graphicx}
\usepackage{subcaption}
\usepackage{booktabs}
\usepackage{tabularx}
\usepackage{array}
\usepackage{ragged2e}
\usepackage{enumitem}

\usepackage{hyperref}
\usepackage{xurl}  

\usepackage[accepted]{icml2026}

\usepackage{amsmath}
\usepackage{amssymb}
\usepackage{mathtools}
\usepackage{amsthm}
\usepackage[capitalize,noabbrev]{cleveref}

\theoremstyle{plain}

\theoremstyle{definition}

\theoremstyle{remark}

\usepackage[disable,textsize=tiny]{todonotes}

\newif\ifshowacknowledgements
\showacknowledgementstrue

\icmltitlerunning{What Should Frontier AI Developers Disclose About Internal Deployments?}

\begin{document}

\twocolumn[
  \icmltitle{What Should Frontier AI Developers Disclose About Internal Deployments?}

  \icmlsetsymbol{equal}{*}

  \begin{icmlauthorlist}
    \icmlauthor{Jacob Charnock}{mats}
    \icmlauthor{Raja Mehta Moreno}{mats}
    \icmlauthor{Justin Miller}{mats}
    \icmlauthor{William L. Anderson}{mats}
  \end{icmlauthorlist}

  \icmlaffiliation{mats}{MATS Research, Berkeley, California, United States of America}

  \icmlcorrespondingauthor{Jacob Charnock}{jakecharnock25@gmail.com}

  \icmlkeywords{AI governance, internal deployment, transparency, model reporting, frontier AI}

  \vskip 0.3in
]

\printAffiliationsAndNotice{}  

\begin{abstract}
  Frontier AI developers are increasingly deploying highly capable models internally to automate AI R\&D, but these deployments currently face limited external oversight. It is essential, therefore, that developers provide evidence that internally deployed models are safe. While recent work has highlighted the risks of internal deployments and proposed broad approaches to transparency and governance, there remains little guidance on the specific information developers should disclose about them. We address this gap by identifying key information that developers should disclose about internally deployed models across four categories: capabilities, usage, safety mitigations, and governance. For each category, we analyse the key benefits and risks of disclosure and consider how disclosure-related risks can be mitigated. Our framework could be used by developers to inform both public transparency documents, such as model system cards, and private periodic reports required under emerging frontier AI regulation such as SB~53, NY~RAISE, and the EU GPAI Code of Practice.
\end{abstract}

\section{Introduction}
\label{sec:introduction}

Frontier AI developers are increasingly relying on highly capable models for machine learning engineering (MLE) and AI research and development (AI R\&D) internal to AI labs. As these deployments grow in scope and capability, it is essential that developers provide evidence to policymakers and the public that the use of these models is sufficiently safe. Internal deployments can pose unacceptable misuse, misalignment, and security risks \citep{ai-models-can-be-dangerous-before-public-deployment, stix2025aicloseddoorsprimer}. Some existing regulatory frameworks touch on the governance of internal deployments. For example, California's SB~53 requires developers to assess and manage catastrophic risk from internal use and to submit quarterly summaries of these assessments \citep{sb53}. Similarly, the EU GPAI Code of Practice requires model reports to contain ``a description of how the model has been used and is expected to be used, including its use in the development, oversight, and/or evaluation of models'' \citep[Measure~7.1]{euGPAICode}. However, neither specifies what information about internal deployments should be disclosed or made public. Moreover, there is no standard procedure for transparently reporting on internally deployed models.

We define internally deployed models (henceforth, \emph{IDMs}) as models that are deployed within labs to conduct privileged tasks. Privileged tasks are tasks that materially shape the developer's AI research, development, or operational infrastructure, namely AI R\&D, machine learning engineering, training data generation, model evaluations, alignment and safety research, and maintenance of core development and deployment infrastructure. This definition deliberately diverges from those in the existing literature---which define IDMs as ``the act of making an AI system available for access and/or usage exclusively for the developing organization'' \citep{stix2025aicloseddoorsprimer, kwon2026internaldeploymentgapsai, ISRSAA2025}---to avoid excluding many important types of internally deployed models that are used for safety-critical tasks internally but may also be available outside the developer.\footnote{For extended discussion on this choice, see Appendix~\ref{app:unreleased}.}

Recent work identifies potential risks posed by IDMs. These include models exploiting privileged access to sensitive infrastructure such as model weights and training pipelines \citep{stix2025aicloseddoorsprimer, acharyaDelaney2025}, autonomously pursuing misaligned or unintended goals \citep{ai-models-can-be-dangerous-before-public-deployment},\footnote{This can include models that pursue undesirable goals through scheming behaviours \citep{stix2025aicloseddoorsprimer}.} recursive self-improvement \citep{field2026airesearchersviewsautomating}, and vulnerability to theft and sabotage \citep{acharyaDelaney2025, ai-models-can-be-dangerous-before-public-deployment}. Moreover, the current lack of transparency into IDMs creates information asymmetries that prevent policymakers from assessing risks \citep{kwon2026internaldeploymentgapsai, acharyaDelaney2025, ai-models-can-be-dangerous-before-public-deployment}. Internal deployments also raise concerns about concentration of power within labs or misuse of highly capable internal systems by lab insiders without external oversight \citep{davidson2025aienabledcoupsho, stix2025aicloseddoorsprimer}.

Despite this urgency, policymakers and the public are ill-equipped to measure or govern the risks from internal deployments due to a continued lack of visibility. Several governance solutions acknowledge the need for greater transparency and have proposed regularly reporting information about IDMs \citep{kwon2026internaldeploymentgapsai, stix2025aicloseddoorsprimer, acharyaDelaney2025}, third-party evaluations or oversight bodies \citep{stix2025aicloseddoorsprimer, acharyaDelaney2025}, and incident reports \citep{kwon2026internaldeploymentgapsai}. Despite some guidance on what information might be useful to disclose about AI R\&D \citep{chan2026measuringairdautomation}, it remains unclear what specific information developers should disclose about IDMs. There also remains uncertainty about the best way to disclose this information and the associated benefits and risks.

This paper aims to fill this gap by identifying the key information that developers should disclose about IDMs across four categories: their capabilities, usage, safety mitigations, and governance. We argue that the lack of transparency into IDMs is significant, and that improving transparency through specific information reporting is crucial to mitigate risks from internal AI applications (e.g., AI R\&D). For each of our proposed information categories, we analyse the key benefits and risks of disclosing that information, and explain how labs could provide this information whilst mitigating security and intellectual property concerns. We finish by discussing how developers could practically implement these disclosures. We argue that the risks of disclosing information about IDMs can likely be mitigated by limiting public disclosures to high-level or redacted summaries, and sharing more sensitive details with trusted third parties and regulators through secure channels commonly used in established auditing contexts. Our proposed disclosures could inform what developers include in both public reporting about IDMs (e.g., in model system cards) and non-public reports to regulators about internal AI use, such as those required under SB~53, NY~RAISE, and the EU GPAI Code of Practice.

\section{Motivating transparency into internal deployment}
\label{sec:motivation}

This section outlines how greater transparency into internal deployments can enable more effective external oversight, improve safety practices, and increase stakeholder trust.

\paragraph{More effective external oversight.} Greater transparency into IDMs could significantly improve the ability of governments, policymakers, and third parties to identify and track emerging AI risks \citep{acharyaDelaney2025, ISRSAA2025, ISRSAA2026}. This is particularly important as models deployed internally for AI R\&D could, if misaligned or compromised, manipulate safety-relevant research, insert exploitable vulnerabilities into code, or poison the training data of successor models \citep{anthropic2026sabotage, acharyaDelaney2025}. Specific disclosures can serve as proxies for risk severity. For instance, benchmark results on evaluations such as RE-Bench or SWE-Bench Pro can allow third parties to track capability jumps in domains where autonomous AI may pose significant risks \citep{pmlr-v267-wijk25a, deng2025swebenchproaiagents, field2026airesearchersviewsautomating}. Additionally, information about model autonomy and human oversight can indicate the potential scale and severity of internal accidents \citep{chan2026measuringairdautomation}. Transparency into internal deployments can also help auditors and evaluators more effectively characterise risks, as major internal deployments that are not publicly announced could contribute disproportionately to a developer's risk profile \citep{brundage2026frontieraiauditingrigorous}. More broadly, transparency into IDMs is likely a prerequisite for effective policy and safety interventions, as policymakers and third parties cannot develop informed recommendations without first understanding how models are used internally.

\paragraph{Improved safety practices.} Greater transparency into IDMs could also improve the safety practices of frontier AI developers. Disclosing information about internal model use and safety practices can encourage developers to adopt responsible practices, increase the extent to which they adhere to them,\footnote{As an example, \citet{10.1108/02686900510570669} find that employees tend to improve their performance when anticipating an audit.} and allow independent stakeholders to identify areas of non-compliance \citep{homewood2025thirdpartycompliancereviewsfrontier}, or vulnerabilities, that developers themselves may overlook. For example, David Rein, a member of METR staff, recently spent three weeks red-teaming a subset of Anthropic's internal agent monitoring and security systems and discovered several specific novel vulnerabilities \citep{red-teaming-anthropic-s-internal-agent-monitoring-systems}. Similarly, Anthropic and OpenAI provided UK AISI and US CAISI with access to non-public versions of classifiers enabling government red-teamers to uncover vulnerabilities including prompt injection vulnerabilities, cipher-based evasion techniques, and universal jailbreaks that were then used to strengthen their safeguards \citep[see][]{anthropic2025caisi, openai2025caisi}. Transparency may also encourage developers to adopt better internal deployment practices from a desire to be viewed as ``best in class'', particularly when competitors publicly demonstrate more responsible internal deployment practices \citep{alaga2024gradingrubricaisafety}.

\paragraph{Greater stakeholder trust.} Transparency into IDMs can also provide greater external assurance to governments and the public that internal deployments are sufficiently safe. Developers that disclose information about IDMs, including their use cases and safety mitigations, can establish greater trust in their internal practices than those that provide no information. This is partly because disclosures create a verifiable record against which future claims and practices can be assessed \citep{brundage2020trustworthyaidevelopmentmechanisms}. Transparency into IDMs also improves visibility into new and emerging risks, enabling both developers and policymakers to make better informed decisions on risk management \citep{Kolt_Anderljung_Barnhart_Brass_Esvelt_Hadfield_Heim_Rodriguez_Sandbrink_Woodside_2024}. Developers also have an incentive to disclose this information insofar as they want to assure policymakers that they are measuring and mitigating systemic and catastrophic risks effectively, as is required through both public AI frameworks and private reports to regulators under frontier AI legislation such as the EU GPAI Code of Practice \citep{euGPAICode}, California's SB~53 \citep{sb53}, and NY~RAISE \citep{raise}. Greater transparency also provides assurance to internal stakeholders, such as board members and members of safety teams \citep{homewood2025thirdpartycompliancereviewsfrontier}, that safety is being taken seriously for the internal use of AI models.

In addition to the benefits of greater transparency, developers may face legitimate concerns when disclosing information about IDMs. For example, revealing details about internal model usage and safeguards could expose proprietary workflows to competitors or provide adversaries and misaligned models with information that could be used to evade safety mitigations. We outline and respond to the potential risks of greater transparency in Section~\ref{sec:what-to-disclose}, and suggest most of these risks can be mitigated using existing practices such as limiting public disclosures to high-level summaries and sharing sensitive information with trusted third parties.

\section{What information should be disclosed about IDMs?}
\label{sec:what-to-disclose}

In this section, we present our example disclosure table (\cref{tab:disclosure}), which outlines the key information that developers should disclose about IDMs across four categories: their capabilities, usage, safety mitigations, and governance. For each category, we outline specific disclosure examples and discuss the key benefits and risks of disclosing this information. We also discuss ways to mitigate potential risks of disclosing this information.

\begin{table*}[p]
\centering
\caption{Example Disclosure Table: Disclosing information about internally deployed models across four key categories.}
\label{tab:disclosure}
\begin{tabularx}{\textwidth}{@{}>{\RaggedRight\arraybackslash}p{0.22\textwidth} >{\RaggedRight\arraybackslash}X@{}}
\toprule
\textbf{Information category} & \textbf{Example disclosures} \\
\midrule
\raisebox{-0.5\baselineskip}[0pt][0pt]{Capabilities} &
\begin{itemize}[leftmargin=1.2em,topsep=0pt,partopsep=0pt,itemsep=2pt,parsep=0pt]
  \item \textbf{How are internal deployments related to external models?} (i.e., are they part of the same model family, fine-tuned, or distilled variants etc.)
  \item \textbf{How do IDMs score on a few public benchmarks you consider especially relevant to automating AI R\&D?} (e.g., RE-Bench, SWE-Bench Pro)
  \item \textbf{What is the proportion of internal inference allocated to IDMs vs.\ non-IDMs?} (measured in FLOPs, GPU-hours, or tokens, reported as a mean taken over the month before publication)
  \item \textbf{Broadly, how do IDMs' capabilities diverge from public-facing models in relevant domains?} (i.e., step changes in key domains such as cyber-offence)
\end{itemize} \\
\addlinespace
\raisebox{-0.5\baselineskip}[0pt][0pt]{Usage} &
\begin{itemize}[leftmargin=1.2em,topsep=0pt,partopsep=0pt,itemsep=2pt,parsep=0pt]
  \item \textbf{On which kinds of tasks are IDMs deployed?} (e.g., coding, research assistance, AI evaluations, AI R\&D, and training data generation)
  \item \textbf{How autonomous are IDMs?} (i.e., are rollouts generally unsteered by humans for seconds, minutes, hours, or days?)
  \item \textbf{What proportion of changes to your frontier model training pipeline(s) is reviewed by a human expert and what reviewing methods are used?}
\end{itemize} \\
\addlinespace
\raisebox{-0.5\baselineskip}[0pt][0pt]{Safety mitigations} &
\begin{itemize}[leftmargin=1.2em,topsep=0pt,partopsep=0pt,itemsep=2pt,parsep=0pt]
  \item \textbf{What safety mitigations are in place for IDMs?} (e.g., output monitoring systems, control mechanisms, human-in-the-loop requirements, interaction logging, and specific monitoring techniques such as probes or classifiers)
  \item \textbf{How have you stress tested your safeguards?} (e.g., have you explicitly tested if you can get the model to consistently fool monitors through prompting, fine-tuning, and similar elicitation techniques?)
  \item \textbf{How do IDMs score on misalignment- and monitorability-related evaluations?} (e.g., evaluation awareness, persuasion, and CoT-monitorability/faithfulness evaluations)
\end{itemize} \\
\addlinespace
\raisebox{-0.5\baselineskip}[0pt][0pt]{Governance} &
\begin{itemize}[leftmargin=1.2em,topsep=0pt,partopsep=0pt,itemsep=2pt,parsep=0pt]
  \item \textbf{Which uses of IDMs are prohibited?} (e.g., autonomous code deployment to production, unsupervised access to customer data, self-modification of model weights, and access to GPU clusters)
  \item \textbf{What are your internal governance mechanisms for employees?} (e.g., staff acceptable use policies, incident reporting and response channels)
  \item \textbf{What concerning model behaviours do you monitor and what are your planned responses to observing concerning behaviour?}
  \item \textbf{How are internal deployment practices reviewed or updated over time?} (e.g., periodic risk reassessments, post-incident reviews, policy revision cadence, and internal/external audits)
  \item \textbf{Who has access to IDMs?} (i.e., approximate number of employees and their level of seniority, tiered access restrictions)
\end{itemize} \\
\bottomrule
\end{tabularx}
\end{table*}

\paragraph{Capabilities.} This category covers key information about the capabilities of IDMs, including how IDMs relate to the developer's publicly available models (e.g., model families), perform on key benchmarks, and their computational resources. The main benefit of disclosing capability information is that it indicates to third parties the potential risk profile of IDMs and the pace of internal AI progress. For example, insight into IDM compute allocation (i.e., the fraction of internal inference that is performed on all IDMs vs.\ public models) could allow third parties to make more accurate predictions about IDM capabilities, given that inference-time compute scaling can amplify a model's effective capabilities far beyond training compute \citep{ord2025inferencescalingreshapesai, deepmind2026deepthink}. Shifts in compute and staff allocation towards internal use can also signal the pace at which AI R\&D is being automated \citep{chan2026measuringairdautomation}. Disclosing whether IDMs are part of the same model family as external models (e.g., distilled or fine-tuned variants) can help third parties infer the alignment and safety training these models likely underwent \citep[e.g.,][]{openai2026gpt5miniNano}. It can also clarify which models are influencing the development of future systems, making it easier to trace capability and behavioural lineage across model generations \citep{cloud2025subliminallearninglanguagemodels}. Finally, tracking internal performance on specific benchmarks (e.g., RE-Bench or SWE-Bench Pro) enables third parties to identify capability jumps in domains where autonomous AI may pose significant risks \citep{pmlr-v267-wijk25a, deng2025swebenchproaiagents, field2026airesearchersviewsautomating}.\footnote{It would be useful to report scores on RE-Bench and SWE-Bench Pro because these benchmarks are increasingly used to forecast progress towards highly autonomous systems \citep[see][]{ai2025forecast}.}

There are some concerns associated with providing information about IDM capabilities. Details about compute allocation and model family, for instance, could reveal sensitive R\&D strategy (e.g., internal resource allocation) and provide competitors with sensitive business information (e.g., knowledge of a new pre-training run well before release). Public disclosures revealing a significant shift in compute towards internal use could also intensify competitive pressure, as competitors may feel compelled to accelerate their own internal AI R\&D in response. Regularly benchmarking IDMs may also be burdensome, as high-complexity agentic evaluations such as RE-Bench and SWE-Bench Pro are currently difficult to run reliably, and require significant compute infrastructure and careful methodological choices \citep{pmlr-v267-wijk25a, deng2025swebenchproaiagents}.\footnote{These benchmarks may also be approaching saturation for the most capable models \citep{anthropic2026opus46SystemCard}, reducing their informational value over time.}

Nevertheless, we think most of these risks can largely be mitigated. To mitigate the risk of revealing sensitive strategy, developers can initially provide model family information to auditors and regulators, and release it publicly after a short delay, as previous model cards have done \citep[e.g.,][]{openai2026gpt5miniNano}.\footnote{For instance, \citet{openai2026gpt5miniNano} disclosed that their mini and nano models are ``smaller'' and ``faster'' variants of their flagship models.} To avoid revealing detailed compute allocation information, developers could disclose normalised or high-level summaries that still enable independent oversight of capability acceleration \citep{chan2026measuringairdautomation}. This mitigation seems feasible given that Anthropic already tracks the pace of model improvements using metrics like ``raw compute scaleup'' \citep{anthropic2026scaleup} and separately revealed internal staff experienced a mean productivity uplift of 4x from using their Mythos Preview model \citep{anthropic2026mythosPreview}. In cases where compute allocation information cannot be summarised (e.g., GPU hours), or may exacerbate race dynamics,\footnote{It should be noted that reporting on compute allocation to internal models can have negative consequences for race dynamics---if, for example, developers suddenly notice increases in competitors' compute allocation into AI R\&D internally, this could cause them to race.} developers should provide disclosures to third parties under NDAs or in non-public reports to regulators. Finally, to address the burden of regularly benchmarking IDMs, developers could default to reporting benchmark scores for major internal model updates, or every 90 days---whichever is sooner. In practice, however, this is unlikely to be too burdensome as reporting performance on RE-Bench and SWE-Bench Pro is already standard practice for frontier model releases \citep[e.g.,][]{anthropic2026opus46SystemCard, openai2026codexCard, deepmind2026geminiCard}, and Anthropic disclosed both detailed benchmark results and a significant leap in cyber capabilities for its unreleased Mythos model before release \citep{anthropic2026mythosPreviewSystemCard}.

\paragraph{Usage.} This category covers key information about how models are used, including details about the tasks they are given (e.g., coding, AI R\&D, generating training data) and the degree of human oversight for these tasks. The main benefit of disclosing information about how models are used is that it can allow third parties to understand the potential risks and ramifications of internal incidents. For example, knowledge of the tasks that IDMs are used to perform can help third parties identify potentially high-risk applications such as models contributing to frontier model training pipelines or generating synthetic training data \citep{acharyaDelaney2025}. Moreover, information about the degree of autonomy IDMs have, such as the proportion of model-generated changes that undergo human review, and the duration models operate unsteered, can indicate the potential scale of risks and demonstrate responsible internal use practices to stakeholders \citep{chan2026measuringairdautomation}. 

Declaring information about IDM usage can present some IP and PR concerns for developers. First, it could indicate proprietary workflows such as where in the training pipeline models are used and for what purposes. It could also signal entirely novel use cases to competitors. Additionally, disclosing that models operate with a high degree of autonomy or limited human oversight could attract public or regulatory scrutiny.

Nevertheless, we think most of these risks are minimal or can be managed effectively by existing practices. To avoid disclosing their proprietary workflows, developers can limit public disclosures to high-level summaries of how models are used (e.g., for coding assistance, to generate synthetic data). This would only slightly expand on public declarations developers already make about IDM use---for instance, OpenAI's disclosure that ``the research team used Codex to monitor and debug the training run for Codex releases'' \citep{openai2026codexBlog}, and that Anthropic's Mythos model was ``widely used internally during the later stages of its development'' \citep{anthropic2026mythosPreview}. Meanwhile, more sensitive information about the specific pipelines models are used within and how much autonomy they have can be disclosed to trusted third parties. Developers could safely disclose this information by expanding arrangements such as Anthropic's collaboration with METR on their Sabotage Risk Report, in which Anthropic shared ``pre-existing internal materials'' and responded to written and verbal questions from METR about the risks posed by their models \citep{metr2025sabotageReview, metr-2026-rd-section-anthropic-risk-report-feb-2026-review}. However, it may not be possible to fully alleviate concerns about reporting use cases that could be harmful for a developer's reputation. Limiting public disclosures to high-level summaries can reduce the risk of public reputational damage, but cannot remove the risk that regulators or third parties find internal practices concerning. Despite this, identifying concerning use cases is ultimately in developers' interest, as it allows them to address risks before they escalate into more serious incidents or enforcement actions.

\paragraph{Safety mitigations.} This category covers key information about the safety mitigations in place for IDMs, including general alignment training measures, monitoring or oversight mechanisms, and the extent to which those safety mitigations have been stress tested for vulnerabilities. The main benefit of disclosing information about the strength of safeguards and specific examples (e.g., output monitoring systems, human-in-the-loop requirements, linear probes) is that it provides external assurance that developers are taking adequate mitigations for IDMs. Additionally, disclosing whether and how these mitigations have been stress tested (e.g., revealing the elicitation efforts developers run to test if their safeguards can be bypassed) provides further external assurance that developers' safety mitigations remain effective. Further, developers should provide scores on misalignment and monitorability evaluations to help ensure the safeguards implemented are appropriate to the observed levels of misalignment risk \citep{williams2025prodevals, guan2025monitoringmonitorability, kutasov2025shadearenaevaluatingsabotagemonitoring}. These results can also provide evidence for safety cases about internal use.

Disclosing information about IDM safeguards and mitigations can also help developers improve their own safety practices as expert third parties use this information to identify potential weaknesses or vulnerabilities \citep[see e.g.,][]{red-teaming-anthropic-s-internal-agent-monitoring-systems, anthropic2025caisi, openai2025caisi}.\footnote{METR have begun developing methodologies specifically for assessing monitor robustness, such as prototype evaluations testing whether agents can discreetly bypass monitoring systems, which may aid them in vulnerability discovery in future engagements of the kind Rein conducted \citep{early-work-on-monitorability-evaluations}.} 

At the same time, declaring information about internal security mitigations and monitoring practices could present security challenges and cause reputational damage for developers. Disclosing specific safeguards (e.g., internal monitoring setups or classifiers), the ways in which developers stress test their classifiers, and the security measures they implement to protect models from theft, could inadvertently expose security vulnerabilities. Developers may be reluctant to disclose information about stress-testing results or identified vulnerabilities if this evidence points towards inadequate safeguards or weak internal safety practices.

Nevertheless, most of these concerns can be managed. Firstly, not all information about safety mitigations carries the same risk from disclosure. Here, we draw a distinction between \emph{alignment} practices, which aim to shape model behaviour \citep{anthropic2026constitution}, and \emph{safeguards}, which aim to catch specific violations by adversarial users or misaligned models. Information on alignment practices (e.g., constitutional AI, RLAIF) will generally not reveal specific vulnerabilities, whilst disclosing specific safeguards (e.g., internal monitoring setups or classifiers) could expose vulnerabilities. To overcome security risks from sharing more security-critical information about safeguards, developers could provide documents to trusted third parties through secure document management and transfer tools (e.g., encrypted file transfers or secure collaboration platforms) \citep{10193389} that are routinely used in auditing contexts \citep{iso27001, fca-sup5-6, occ2019}. Public disclosures can also be limited to redacted materials or high-level accounts of safeguards to preserve sensitive details \citep{cfr2026}. OpenAI, for instance, has described how it monitors internal coding agents for signs of misalignment \citep{openai2026codingMonitor}. Despite this, developers may still be reluctant to disclose information about safety mitigations if this evidence points towards weak internal safety practices. As noted above, sharing findings with regulators can limit public reputational damage, but cannot prevent regulators from finding internal practices inadequate. However, any residual risk is likely outweighed by the benefits of identifying weaknesses early, and the fact that SB~53 and NY~RAISE will require disclosures about internal risk mitigations \citep{sb53, raise}.\footnote{SB~53 \S~22757.12(a) requires large frontier developers to publish a frontier AI framework describing how they approach, among other things, ``applying mitigations to address the potential for catastrophic risks'' (a)(3), ``reviewing assessments and adequacy of mitigations as part of the decision to deploy a frontier model or use it extensively internally'' (a)(4), and ``assessing and managing catastrophic risk resulting from the internal use of its frontier models, including risks resulting from a frontier model circumventing oversight mechanisms'' (a)(10). Similarly, NY~RAISE \S~1421(1)(j) requires frameworks to address ``catastrophic risk resulting from the internal use of its frontier models, including risks resulting from a frontier model circumventing oversight mechanisms''; \S~1422(2)(a) mandates confidential quarterly internal-use risk summaries.}

\paragraph{Governance.} This category covers key information about developers' internal governance of IDMs, including the prohibited uses for IDMs, acceptable use policies, and whether and how governance practices are reviewed and updated over time. The main benefit of disclosing governance information is that it provides assurance to stakeholders that internal deployments are governed responsibly. It also enables policymakers and third parties to assess whether internal risk management systems are adequate. For example, disclosing prohibited uses allows external parties to assess whether developers have considered key threat models such as self-modification of model weights or unsupervised access to GPU clusters \citep{shlegeris2025}. 

Governance disclosures can also reveal the concrete steps developers take to mitigate risks (e.g., 24-hour alignment stress-testing window before wide internal deployment of Claude Mythos Preview \citep{anthropic2026mythosPreview}), helping external parties gauge the maturity of internal practices and surface potential gaps. Disclosing who has access to IDMs can help third parties assess insider risk and the concentration of control over powerful internal systems \citep{davidson2025aienabledcoupsho, stix2025aicloseddoorsprimer, what-should-companies-share-about-risks-from-frontier-ai-models}. Transparency into governance mechanisms for staff (e.g., acceptable use policies, access restrictions, and incident reporting channels/whistleblower protections) can also increase accountability by confirming that risks can be flagged by employees through designated channels \citep{openai2026raising, what-should-companies-share-about-risks-from-frontier-ai-models}. Disclosing information about recent incidents involving internal deployments (e.g., egregious reward hacking, attempted rogue deployments, or attempted self-exfiltration) can also help third parties assess the frequency and severity of risks that materialise in practice \citep{chan2026measuringairdautomation}. Finally, it is important to disclose how often governance policies are reviewed and updated given that IDMs will evolve continuously \citep{kwon2026internaldeploymentgapsai}, meaning governance practices must be updated at a proportionate cadence.

Despite these benefits, information about internal governance presents some potential risks. For one, developers may be reluctant to disclose governance practice if this reveals inadequate internal processes (e.g., insufficient safeguards, or no whistleblower protections) that could cause reputational damage. Relatedly, pressure to publicly disclose information about internal governance could discourage developers from scrutinising their own internal practices too closely, to avoid having to disclose unfavourable findings \citep{what-should-companies-share-about-risks-from-frontier-ai-models}. There is also a risk that governance disclosures could enable ``marginal risk arguments'', where developers point to competitors with weaker internal governance to justify that their own practices are sufficient and do not need to be improved \citep{williams2025}.

However, most of these risks are manageable or unlikely to be significant in practice. For example, OpenAI describes its internal monitoring and escalation processes including the functions of its Safety Oversight and Security Incident Response teams \citep{openai2026codingMonitor}, as well as their ``Raising Concerns Policy'', which outlines how employees can raise issues \citep{openai2026raising}. These examples suggest that public disclosure of some governance information is feasible. To avoid disclosing more sensitive details about internal governance structures including which employees can access internal models, and details of internal/external audits, developers can limit public disclosure to high-level summaries and provide more detailed, sensitive information with trusted third parties and regulators through secure channels routinely used in auditing contexts \citep{10193389, iso27001, fca-sup5-6, occ2019}. 

Developers may still be reluctant to disclose governance practices that reveal inadequate internal processes. In such cases, third parties can agree to keep reviews confidential to reduce reputational risks, which can also make developers less averse to scrutiny and lead to more open and collaborative reviews \citep{homewood2025thirdpartycompliancereviewsfrontier}. Given that disclosing information about internal governance is already required under commitment 8 of the EU GPAI Code of Practice \citep{euGPAICode}, and that incident reporting is separately required under both SB~53 and the EU GPAI Code of Practice,\footnote{SB~53 (\S~22757.13) requires reporting critical safety incidents within 15 days, or within 24 hours if the incident poses an imminent risk of death or serious physical injury. Measure 9.3 of the EU GPAI Code of Practice requires reporting serious incidents within 2 to 15 days, depending on severity.} residual reputational risks are unlikely to outweigh the need for providing this information.

\section{Discussion}
\label{sec:discussion}

In this section, we discuss details about how developers could practically provide our suggested disclosures. We address the main limitations of our work and directions for future research in Section~\ref{sec:limitations}.

\paragraph{Where should information be reported?}
Our primary focus in this paper was to outline what information should be disclosed about IDMs, rather than where this information should be disclosed. Nonetheless, two reporting approaches seem most natural: (1) incorporating a dedicated section on IDM use within existing model system cards, or (2) publishing a standalone document that periodically reports on internal deployment practices independently of specific model releases (e.g., every 3 months). Both approaches have separate benefits and risks. Reporting on IDM use within model system cards is likely more convenient for developers, as it does not require creating a new reporting document and can narrowly scope disclosures to how IDMs contributed to the development of the released model. However, tying reporting to model release cycles means IDM use cases that are not directly linked to the production of a particular model (e.g., ongoing AI R\&D, AI evaluations) may go unreported. 

A standalone periodic report, by contrast, would require reporting across all internal deployment practices on a regular schedule regardless of release timelines---thus providing a more complete picture of internally deployed models. It also does not depend on developers continuing to publish model system cards for newly released models. However, it does require developers to establish a new reporting practice. We believe that regular ongoing reports, rather than updates that accompany closed-source model releases, best capture the ongoing nature of risks from internal AI deployments, and neatly map to emerging regulations that require periodic reports on how developers assess catastrophic risk from internal models.

\paragraph{What information is most important to report?}

We chose the questions in \cref{tab:disclosure} carefully, and believe disclosures aimed at addressing them all would meaningfully advance the state of government and public knowledge about IDMs. However, we recognise that alongside these benefits, companies would incur legitimate trade-offs. To accommodate this fact, we suggest prioritising the following four disclosures from our table that we believe offer significant benefit whilst presenting minimal burden and risks to companies:
\begin{enumerate}
    \item \textbf{What concerning model behaviours do you monitor and what are your planned responses to observing concerning behaviour?}
    \item \textbf{Broadly, how do IDMs' capabilities diverge from public-facing models in relevant domains?}
    \item \textbf{What safety mitigations are in place for IDMs?}
    \item \textbf{What proportion of changes to your frontier model training pipeline(s) is reviewed by a human expert and what reviewing methods are used?}
\end{enumerate}

Please see Appendix~\ref{app:prioritised-disclosures} for our full justification for this selection. We encourage future work that refines and expands our suggestions as AI progress continues.

\paragraph{To whom should information be reported?}
As we outlined in Section~\ref{sec:what-to-disclose}, some disclosures about IDMs present security, IP, and PR risk for developers. We reviewed the benefits and risks of providing more information, and argued that key challenges can likely be mitigated through providing high-level public summaries and sharing confidential information with third parties and regulators through established auditing practices used in other industries. In this case, we think there are two clear avenues for disclosure. First, there should be a high-level public report that discloses key information about IDMs and, in some cases, redacts sensitive details that even at a high level cannot be disclosed publicly. Second, developers can share more detailed information in non-public reports to trusted third parties such as independent evaluators (e.g., METR) or regulators (e.g., CA Office of Emergency Services, The EU AI Office). 

Until there is further clarity about the risk of disclosing different information, it will be up to labs to determine the information they disclose publicly vs.\ privately. Nevertheless, we hope our disclosure table will improve IDM reporting by (i) clearly specifying which information developers could include when reporting on IDM use and (ii) helping policymakers and regulators prioritise which information they should request from developers as part of emerging frontier AI legislation. Future work should more clearly articulate which disclosures should be provided to the public and to regulators.

\paragraph{How often should these reports occur?}
Determining an appropriate reporting cadence for IDMs is challenging. This is because internal systems can evolve continuously through additional training, integration, and redeployment in ways that external product releases generally do not \citep{kwon2026internaldeploymentgapsai}. This means there are likely no discrete moments where it is natural to report on IDMs. Importantly, reporting should be done frequently enough to capture meaningful changes in internal model usage, without imposing unnecessary burdens on developers. Reporting on a fixed schedule, such as the quarterly cadence already required by SB~53 for catastrophic risk assessments, is likely simpler to enforce and easier to verify \citep[\S~22757.12(d)]{sb53}. By contrast, updating reports based on significant changes or triggers (e.g., capabilities or use) could be more informative, but could strain internal capacity and relies on developers to self-monitor. We propose reporting whenever there is a significant change in a developer's use of IDMs or every 90 days, whichever is sooner.

Future work should develop more concrete triggers for what constitutes ``significant change''. One approach that has recently been proposed is for developers to update their reports on internal use whenever something changes that invalidates their previously published safety case \citep{delaney2026risk}. This could occur, for example, if the capability profile or opportunity for a model to subvert safety mechanisms changes and undermines the previous safety case. \citet{delaney2026risk} further propose that a report's depth should scale with the increase in capability or risk since the last report. So understood, a major capability jump would warrant a comprehensive report, whereas a quarterly report for a system whose risk profile is largely unchanged can mostly reaffirm the prior safety case. This eases some of the compliance burden of regular reporting, sparing developers from filing new reports when nothing significant has changed, whilst still ensuring materially new risks are adequately addressed. New use cases for internal AI models on privileged tasks should still be reported, even if developers think the overall risk level is covered by the previous safety case. This helps to avoid leaving significant use cases unrecorded.

Tying reports to material changes in risk and new privileged use activities also avoids the question of when reporting should occur in the internal deployment model lifecycle. One downside of this requirement, however, is that such changes may be quite frequent, and could create corporate incentives for developers to rush or improperly report their internal use practices.

\paragraph{Why these information disclosures and not others?}
To determine what information should be disclosed about IDMs, we drew on existing literature on internal AI R\&D \citep{chan2026measuringairdautomation, field2026airesearchersviewsautomating, toner2026whenaibuildsai, eth2025willairdautomati} and internal deployments \citep{ai-models-can-be-dangerous-before-public-deployment, stix2025aicloseddoorsprimer, acharyaDelaney2025, kwon2026internaldeploymentgapsai} to identify the types of information most relevant to external oversight and risk mitigation. We organised our disclosures into four categories---capabilities, usage, safety mitigations, and governance---based on a combination of what developers already disclose about frontier models in public system cards (e.g., capability evaluations, model uses, safeguards, and internal governance mechanisms), the specific risks identified in the literature on IDMs and internal AI R\&D, and the information we believe would help alleviate those risks. Before settling on our final table, we considered a wider set of potential disclosures. For example, we discussed whether developers should disclose the total number of internal deployments active at any one time, the duration each has been deployed, and the time horizons of tasks they are currently used for. However, we determined these to be lower priority than disclosures such as what tasks IDMs are used for, how much autonomy IDMs have when completing those tasks, and how those IDMs are monitored.

\section{Limitations and future work}
\label{sec:limitations}

This work has several limitations that should be considered. 

One limitation of focusing on reporting alone is that the contents of a report may be difficult to verify. Without access to internal models and the infrastructure surrounding internal AI use, it can be hard to confirm the accuracy of a report beyond what a developer chooses to disclose. A developer could, for instance, understate the risks posed by its internal models if no external party is able to check what it reports about their capabilities, usage, safety mitigations, and governance. This is especially pressing for internal deployments, which are by default less transparent than external models that can be evaluated by third parties pre-deployment and are subject to public scrutiny post-deployment. Future work should therefore focus on strengthening external oversight. As an example, since this paper was submitted, METR released a risk report where Anthropic, Google, Meta, and OpenAI each gave METR access to their most capable internal models for the first cross-company assessment of internal-deployment risk \citep{metr-2026-frontier-risk-report}. Such engagements can ensure greater external oversight into internal deployments and help verify company claims about the safety of internal AI use. 

Nevertheless, we think developers face only weak incentives to mislead in their reports. While SB~53 does not mandate third-party evaluation directly, developers are still required to ``clearly and conspicuously publish on [their] internet website a frontier AI framework,'' which must describe how they assess and mitigate risks from internal AI use \citep{sb53}. These reports form a written record that can aid regulators and stakeholders in engaging with developers in the event that their safety claims are later found to be incomplete, inaccurate, or misleading.

Our analysis does not cover the time, staff, and resource costs that reporting on IDMs would impose on developers. We expect these costs to be concentrated in the initial setup (i.e., constructing and reconciling an inventory of IDMs and establishing reporting practices). After setup, the marginal cost per report should be comparatively low for developers that already produce model cards, evaluation results, and safety cases, given that our proposed disclosures are designed to draw on such materials. For public disclosures, developers can further reduce the reporting burden by drawing on the confidential quarterly reports on catastrophic risk from internal AI use that they are already required to produce under SB~53 and the NY~RAISE Act, redacting sensitive contents as appropriate.

Relatedly, we do not prescribe which internal function should administer reporting (i.e., a legal team, technical staff, or a dedicated team), as developers differ too widely in structure for any single assignment to be appropriate. Future work should analyse these implementation costs in more detail and study how reporting can be standardised between developers (e.g., through templates) without mandating overly stipulative guidelines.

Our proposal of useful information to disclose about IDMs is non-exhaustive, and will likely need to be updated over time to include additional important information. As we learn more about IDMs, and their capabilities and usage expand, we anticipate gaining more clarity about which information developers should prioritise reporting as well as the appropriate format this reporting might take. We encourage future researchers to iterate on our proposed list.

\section{Conclusion}
\label{sec:conclusion}

In this paper, we have proposed the key information that developers could disclose about internally deployed models. Our suggestions cover four categories of information about internal models: their capabilities, usage, safety mitigations, and governance. We reviewed the benefits and risks of providing this information, and argued that key challenges can likely be mitigated through a combination of providing high-level public disclosures, redacting public reports, or adopting secure auditing practices from other industries. We further discussed practical implementation considerations, including where this information should be disclosed, to whom, and at what cadence. We hope these proposals provide a useful starting point for researchers, members of safety teams, and regulators to use and iterate on, so transparency into IDMs can improve over time.

\section*{Impact Statement}

This paper proposes a framework for what frontier AI developers should disclose about internally deployed models. Its intended societal impact is to strengthen external oversight of internally deployed AI models (including for AI R\&D), reduce information asymmetries between developers and the public, and help inform a set of reporting requirements for internal deployments under legislation such as SB~53, NY~RAISE, and the EU GPAI Code of Practice. We do not anticipate this paper to have significant risks. Nevertheless, we acknowledge that any paper arguing for greater standards could be treated by developers as a ceiling rather than a floor. We mitigate this risk by emphasising that the information we propose developers disclose is non-exhaustive and should be improved upon in future work. We also appreciate that greater transparency into AI R\&D could increase race dynamics, but we think this risk is minimal given that greater transparency for regulators and third parties ensures more effective external oversight of developers' internal safety practices (including for how they use internally deployed models). On balance, we see many benefits from greater transparency into internally deployed models and believe our paper is an important contribution towards this goal.

\ifshowacknowledgements
\section*{Acknowledgements}
Dr.\ Francis Rhys Ward, Yulia Volkova, Aidan Homewood, Michael Chen, Jason Green-Lowe, Joe Kwon, Ben Hayum, Iskandar Haykel, and Titus Buckworth.
\fi

\section*{LLM Use Declaration}

Generative AI tools were used during manuscript preparation. The authors used those tools solely for literature gathering, structuring tables, grammar editing, and style. No tool was used to generate text for the manuscript.

\bibliographystyle{icml2026}
\bibliography{internal_deployments}

\newpage
\appendix
\onecolumn

\section{Internally Deployed vs.\ Unreleased Models}
\label{app:unreleased}

The literature has previously scoped IDMs as models available for access and/or usage \emph{exclusively} by their developers, e.g., by \citet{stix2025aicloseddoorsprimer} and \citet{kwon2026internaldeploymentgapsai}, who define internal deployment as ``the act of making an AI system available for access and/or usage exclusively for the developing organization.'' We disagree with this definition, and believe it may exclude many important types of internally deployed models.

We prefer to define IDMs as models that are deployed within labs to conduct privileged tasks. Privileged tasks are tasks that materially shape the developer's AI research, development, or operational infrastructure. Namely, AI R\&D, machine learning engineering, training data generation, model evaluations, alignment and safety research, and maintenance of core development and deployment infrastructure.

For most concerns, it is unimportant whether a model which is used internally is also used externally. The alignment assessments, control protocols, internal governance mechanisms, and extent of AI R\&D automation are all of major interest, regardless of whether the model being used internally is also available publicly. The public, third parties, and governments have strong interest in understanding how much autonomy models are given internally and what controls are in place to prevent incidents.

There do exist certain unique risks from models available for access \emph{exclusively} or \emph{near-exclusively} to the developing organisation. We prefer the literature to discuss such models as \textbf{unreleased models}, and believe they pose related but distinct risks, e.g., it may be especially difficult to assess the capabilities of the absolute frontier of unreleased models, and consequently to assess the catastrophic misuse risks if these models were stolen by adversaries. These risks are significant and ought to be studied, but remain distinct from the precise considerations around internal deployments. Such models, when actually used internally, are also well-covered by our definition, as models available exclusively for the developing organisation are very likely to be used for privileged tasks and would therefore be subject to the same reporting considerations.

\section{Justification for Prioritised Disclosures}
\label{app:prioritised-disclosures}

We set out the disclosures we would prioritise below, with a case for each.

\paragraph{What concerning model behaviours do you monitor and what are your planned responses to observing concerning behaviour?}
Under the EU GPAI Code of Practice, signatories must already report serious incidents across the full ``model lifecycle'' \citep{euGPAICode}, which maps closely to this disclosure requirement. Given this information would likely need to be covered in reports to the AI Office, the marginal cost of sharing it privately with Cal OES or New York's Department of Financial Services is low.

\paragraph{Broadly, how do IDMs' capabilities diverge from public-facing models in relevant domains?}
This seems reasonable to disclose, as Anthropic's Responsible Scaling Policy (Version 3.0) states that its Risk Reports will cover internally deployed models that may pose risks beyond their public ones \citep{anthropic2026rsp}, and OpenAI's Preparedness Framework applies to any internal agentic system that represents a substantial increase in the capability frontier \citep{openai2025preparedness}. Information explaining how these models diverge may therefore already be required.

\paragraph{What safety mitigations are in place for IDMs?}
Developers are already beginning to describe their safety mitigations publicly. OpenAI documents how it monitors its internal coding agents for misaligned behaviour \citep{openai2026codingMonitor}, and Anthropic has committed to centralised logs that surface concerning behaviour and security threats \citep{anthropic2026roadmap}. Furthering these practices in more detail for IDMs specifically would give third parties and governments greater assurance about internal use and help identify where mitigations could improve.

\paragraph{What proportion of changes to your frontier model training pipeline(s) is reviewed by a human expert and what reviewing methods are used?}
IDMs are increasingly relied upon for privileged tasks where undetected errors or sabotage in AI-generated outputs could propagate into successor models or compromise company infrastructure \citep{acharyaDelaney2025}. Information about the share of such changes that remains under human expertise, as well as the kind of review used, such as manual or AI-assisted, could be a useful empirical indicator for determining potential oversight gaps. There is also minimal cost to providing details about the scale of review and methods used. In future, a more expansive version of this measure could also track the defects found. As \citet{chan2026measuringairdautomation} note, however, such data could reveal oversight gaps and raise safety concerns. It may therefore be more appropriate to keep in private summaries to regulators.

\end{document}